\documentclass[runningheads]{article}

\usepackage{arxiv}

\usepackage[utf8]{inputenc} 
\usepackage[T1]{fontenc}    
\usepackage{hyperref}       
\usepackage{url}            
\usepackage{booktabs}       
\usepackage{amsfonts}       
\usepackage{nicefrac}       
\usepackage{microtype}      
\usepackage{lipsum}
\usepackage{graphicx}
\usepackage[utf8]{inputenc}
\usepackage{caption}
\usepackage{float}
\usepackage{refstyle}
\usepackage[export]{adjustbox}
\usepackage{amsmath}
\usepackage{ulem}
\usepackage{fix-cm}
\usepackage{array}
\usepackage{caption}
\usepackage{subcaption}
\usepackage{longtable}
\usepackage{changepage}
\usepackage{textcomp}
\usepackage{multirow}
\usepackage{fancyhdr}
\usepackage{comment}
\graphicspath{ {./images/} }

\begin{document}

\title{A Novel Multimodal Music Genre Classifier using Hierarchical Attentionand Convolutional Neural Network}


\author{
 Abhilash Nandy \\
  Indian Institute of Technology Kharagpur\\
  West Bengal\\
  India \\
  \texttt{raj12345@iitkgp.ac.in} \\
  \And
 Manish Agrawal \\
  Indian Institute of Technology Kharagpur\\
  West Bengal\\
  India \\
  \texttt{manishagrawal.iitkgp@gmail.com} \\
}
\maketitle
\begin{abstract}
  Music genre classification is one of the trending topics in regards to the current Music Information Retrieval (MIR) Research. Since, the dependency of genre is not only limited to the audio profile, we also make use of textual content provided as lyrics of the corresponding song. We implemented a CNN based feature extractor for spectrograms in order to incorporate the acoustic features and a Hierarchical Attention Network  based feature extractor for lyrics. We then go on to classify the music track based upon the resulting fused feature vector.
\end{abstract}


\section{Introduction} \label{introduction}

Genre classification of music tracks is a well known researched task in the area of music research community. In the field of Machine Learning, making classification is one of the primary objective for which numerous solution methods are proposed. Varying problem domain, challenges the research community to design and optimize their methods. Specially in music, it is not necessarily important for a genre to have tracks with similar acoustic features. Varied collection of tracks in a single genre makes it difficult for any machine learning model to classify tracks.

Generally, music tracks are released having the information of artist and particular genre the song belongs to. Tracks of same genre are written with same textual characteristics. Rap songs usually begin with chorus followed by composing verse and finally words are added in sync with verse. Pop songs starts with intro following verse, pre-chorus, chorus and outro. These pattern of similarities between the writing pattern is leveraged in our proposed model by using the application of natural language processing. We also make use of acoustic characteristics for classifying a song into its genre. The dataset we will be using provides 30s of raw audio mp3 clip. Leveraging the Python package Librosa  we are extracting the pictorial feature representation of acoustic characteristics. 

In our approach, we extract two different kind of features which shall be used by two different learning models to make a fused vector representation of the track. A neural network based classifier will make final classification over the concatenated vector. 

This paper is further organized as follows: Section \ref{rel} provides details about related work along with their findings, Section \ref{data} describes the dataset which has been used, Section~\ref{feat} explains the preprocessing and feature extraction steps used in our model, Section~\ref{prop_soln} gives a detailed description of our model, Section~\ref{exp} provides details of implementing model, and Section~\ref{res} compares our model with known music genre classifiers.

\section{Related Works}
\label{rel}
Music genre classification is a famous problem domain where unique approaches exists as a solution methodology. Depending upon some factors like feature extraction methods, classification algorithms and analytical study for different benchmark datasets, varying range of significant contribution exists. 

One of the most recent and challenging datasets having music genre annotations is the FMA (Free Music Archive) Dataset \cite{defferrard2016fma}. This dataset is challenging, since, it has an imbalanced data, with a lot of variation in the number of tracks per class over all classes. Works on this dataset have varied approaches to solve the classification task at hand. For instance, \cite{kim2018transfer}  learns artist-related information by employing various transfer learning tasks on track metadata and track audio. \cite{murauer2018detecting} uses various boosting, bagging and neural network models in order to classify genres using acoustic and spatio-temporal features extracted from the audio clips. 

\cite{franklin2006recurrent} uses long short-term memory (LSTM) cells for extracting high level features which can be further used for various purposes. \cite{li2010automatic} make usage of CNN models to extract features from raw audio files of song tracks to make predictions over variety of tasks. \cite{tsaptsinos2017lyrics} employs hierarchical attention networks on the set of lyrics dataset to classify the music genre of that song. \cite{dong2018convolutional} combines knowledge of human perception in music genre and neurophysiology of auditory system by employing CNN models across segments of music signals. Moreover, combining the CNN models with RNN has shown some improvements \cite{zuo2015convolutional}. 

Inspired from the works described in the previous paragraph, we decided to employ a fusion of two approaches - one that uses audio clips to make use of spatio-temporal characteristics, and the other uses song lyrics, and applies natural language processing over it.

\section{Dataset Description}
\label{data}
The FMA (Full Music Archive) \cite{defferrard2016fma} dataset comes in four sizes - full, large, medium and small. We make use of the medium version of the dataset. The medium version comprises of 25,000 audio clips of length 30s each with music genre annotations, along with track metadata such as track title, name of the artist etc. The data is spread over 16 genres, with the number of tracks per genre being very imbalanced, varying from 21 to 7,103.

\section{Feature Extraction and Pre-Processing}
\label{feat}
The FMA (Full Music Archive) \cite{defferrard2016fma} dataset comes in four sizes - full, large, medium and small. We make use of the medium version of the dataset. The medium version comprises of 25,000 audio clips of length 30s each with music genre annotations, along with track metadata such as track title, name of the artist etc. The data is spread over 16 genres, with the number of tracks per genre being very imbalanced, varying from 21 to 7,103.

\section{Proposed Solution}
\label{prop_soln}
\begin{figure}[H]
    \centering
    \includegraphics[scale=0.25]{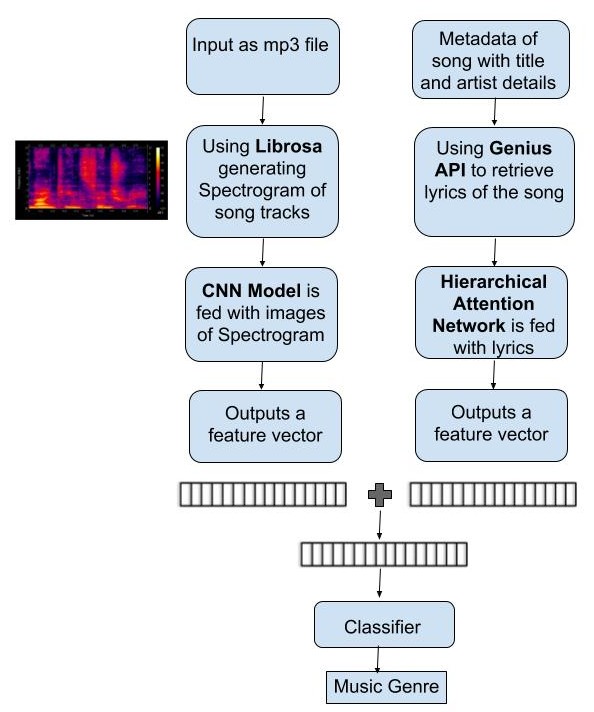}
    \caption{Our Proposed Solution}
    \label{fig:overview}
\end{figure}

In this paper, we focus on a multimodal structure to leverage the advantages of multiple data types. Specifically, given an mp3 file, we retrieve lyrics to use it as the linguistic content and generate spectrogram to use as pictorial feature. Our model comprises of three components, hierarchical attention module over lyrics, convolutional neural network module for spectrogram and finally a feature fusion module which maps the input to its music genre (as shown in Fig. \ref{fig:overview}).

\subsection{Convolutional Neural Network (CNN) for Spectrogram}

\begin{figure}[H]
    \centering
    \includegraphics[width=0.5\textwidth]{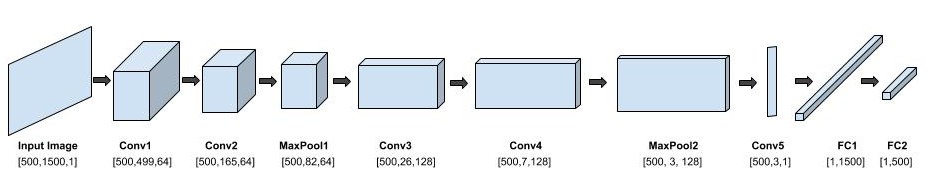}
    \caption{Architecture of the Convolutional Neural Network}
    \label{fig:cnn}
\end{figure}

For extracting features from the tracks represented as spectrograms, a deep Convolutional Neural Network \cite{krizhevsky2012imagenet} based architecture was adapted from work of \cite{murauer2018detecting} on genre detection. The architecture of the CNN network is shown in Fig. \ref{fig:cnn}. In order to capture the temporal variation of amplitudes at different frequencies, a 2D convolution with 1D kernel is incorporated which convolves across the time dimension. Given that the dimension of input is $500$ x $1500$ pixels, the output generated by the CNN model is a feature vector of length $500$.
 
The CNN model uses blocks comprising the convolution layer, batch normalization layer, activation layer and max pooling layer if required. We will be using ReLu as the activation function before a Batch Normalization layer since it is known to speed up the convergence of loss function. We also employ a random dropout after the pooling layer for regularization purpose. 

\subsection{Hierarchical Attention Networks (HANs)}

Our intention is to derive meaning of a sentence from the words and then derive the meaning of song lyrics from those sentences. Thus, we make use of Hierarchical Attention Network (HAN) which is inspired from the work of \cite{yang2016hierarchical}. In this network each layer is fed to a bidirectional gated recurrent unit (GRU) \cite{chung2014empirical} with attention applied to the output. Some sentences of song carry more meaning than others. Attention networks learns weights proportional to their meaning. The attention weights are used to create a vector via a weighted sum which is then passed as the input to the next layer.

\begin{figure}[H]
    \centering
    \includegraphics[width=\linewidth]{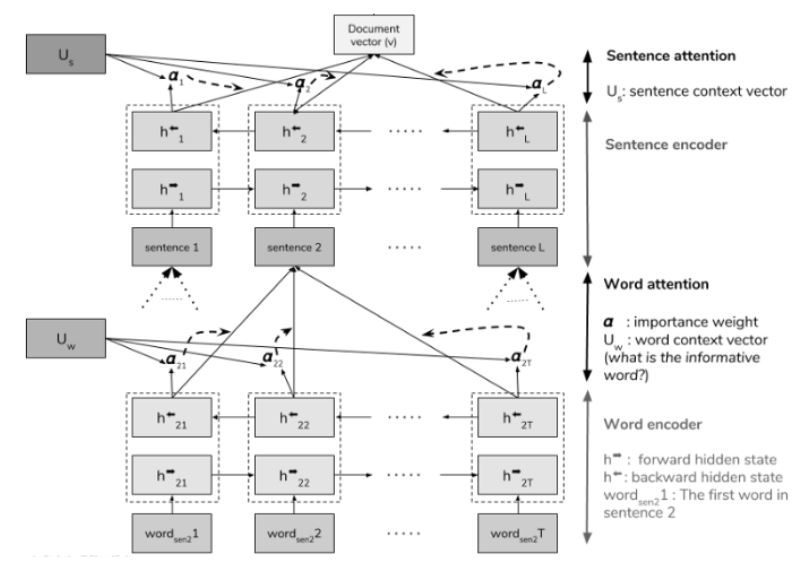}
    \caption{Architecture of the Hierarchical Attention 
    Network}
    \label{fig:HAN}
\end{figure}

A representation of the architecture can be seen in Fig. \ref{fig:HAN}  where the layers are applied at the word, and sentence level. 

\subsubsection{Word Embeddings}
Dense vectors are used to represent words in common NLP applications. Since similar words have similar context, vectors can be learned through their context. Models like word2vec model \cite{word2vec}. GloVe model \cite{glove} are used for representing linguistic contents as vectors.

In our implementation, we first embedded the words into 300-dimensional vectors by word2vec, which gives us the best result compared to GloVe. Unknown words were randomly initialized. Given a sentence S with N words, let $w_i$ represent the $i_{th}$ word. We embed the words through the word2vec embedding matrix $W_e$ by : 
\begin{equation}
    T_i = W_ew_i, i \in [1, N]
\end{equation}
where $T_i$ is the embedded word vector.

\subsubsection{Hierarchical Attention}
\begin{figure}[H]
    \centering
    \includegraphics[scale=0.3]{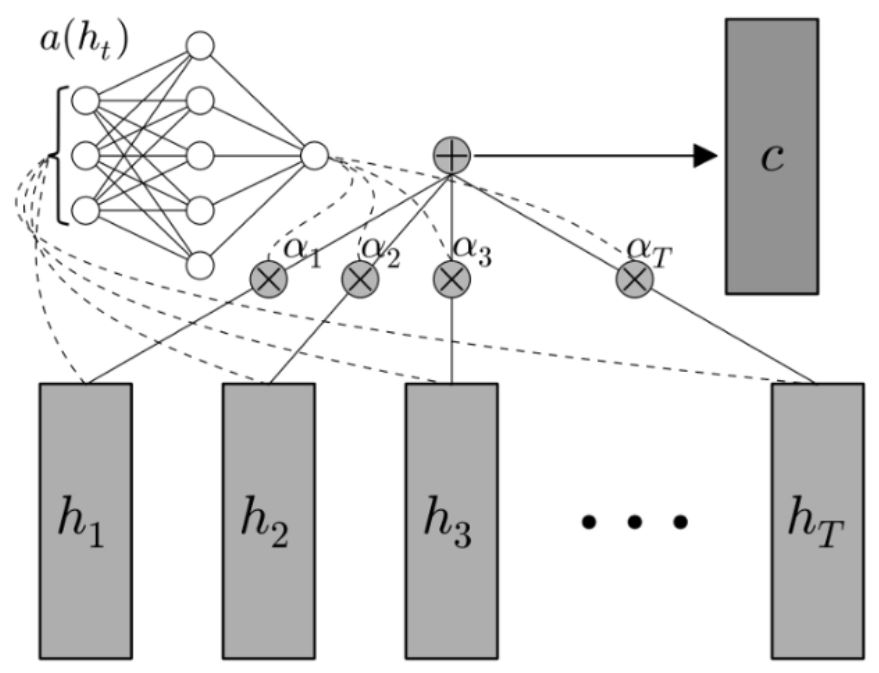}
    \caption{\textbf{Attention Mechanism}: Vectors belonging to hidden state $h_t$ are fed into the trainable function $a(h_t)$ to produce a probability vector $\alpha$. 
    The vector $c$ is computed as a weighted average of $h_t$, with weighting given by $\alpha$}
    \label{fig:attention}
\end{figure}

In order to learn which words (or sentences, as is the case) were more important in the classification objective, a HAN is used, as shown in Fig. \ref{fig:attention}. Along the lines of that study, we would like our model to learn which words are important in classifying genre and then apply more weight to these words. Attention was first proposed by \cite{bahdanau2014neural} for the application of neural machine translation to allow the model to learn the the significance of words in translation purpose.

Given input vectors $h_i$ for $i = 1, . . . , n$ the attention mechanism can be formulated as -
\begin{equation}
    u_i = tanh(W_ah_i + b_a)
\end{equation}
\begin{equation}
    \alpha_i = \frac{exp(u_i^Tu_a)}{\sum_{k=1}^{n} exp(u_k^Tu_a)}
\end{equation}
\begin{equation}
    s = \sum_{i=1}^{n}\alpha_ih_i
\end{equation}

where $s$ is the output vector consisting of the weighted sum of the vectors of current layers. $s$ is passed to the next layer. Parameters $W_a$, $b_a$, and $u_a$ are learned by the model after random initialisation.

All the songs in the training set are padded in such a manner that the number of sentences reaches the maximum number of sentences among all the songs, and each sentence reaches the maximum number of words per sentence among all songs.  Given the lyrics of the song, a bidirectional GRU is applied on the words, followed by attention mechanism to form a weighted sum of hidden states. This is passed through another bidirectional GRU with attention mechanism, which gives an output vector to be used in the fusion stage.

\subsection{Classification}
The feature vectors obtained as outputs from the CNN and the HAN are concatenated. The resulting vector is then fed into a fully connected layer, with number of outputs equal to the number of music genre. Softmax activation function is applied over the outputs to obtain probabilities of association of lyrics with all the classes.
\begin{equation}
    p = softmax(W_ps + b_p)
\end{equation}
In order to train the whole network, categorical cross-entropy loss function is minimised. Cross-entropy loss function is given as :
\begin{equation}
    J = -\sum_{k=1}^{K}log(p_{d_k}k)
\end{equation}
where, $d_k$ is the true genre label of the song.

\section{Experiments}
\label{exp}
A 80/10/10\% split is used for training, validation and test sets. The data is split in a stratified fashion, and artist filters are applied, as proposed in FMA \cite{defferrard2016fma}. Further for comparing our models, training set of 25,000 clips and a testing set of 35,000 clips was also used.

\textbf{Training parameters and hyperparameters} - For the CNN, the dropout parameter is taken to be 0.5. For the HAN, the number of hidden units is chosen to be 50, maximum sentence length is chosen to be 20 and maximum number of sentences in a song is chosen to be 50. SGD optimizer is used along with a Nesterov Accelerated Momentum of 0.9 for training, with a batch size of 32. Convergence is reached after around 200 epochs. 

\section{Results}
\label{res}
The evaluation metrics used in order to compare our model with other models are F1-score, Logloss and test accuracies. Evaluation metric results for various models have been tabulated in Table ~\ref{tab:results}.

\begin{table}[thpb]

\caption{Comparison of results}
\label{tab:results}
\centering
\begin{tabular}{|c|c|c|c|}
\hline
\textbf{Model}                                                & \textbf{F1 score} & \textbf{Logloss} & \textbf{\begin{tabular}[c]{@{}c@{}}Test \\ Acc.\end{tabular}} \\ \hline
\begin{tabular}[c]{@{}c@{}}Kim et al.\\  (MTN)\end{tabular}   & 0.6571            & 0.7727           & --                                                            \\ \hline
\begin{tabular}[c]{@{}c@{}}Murauer and \\Specht\end{tabular}                                               & \textbf{0.78}     & 0.85             & --                                                            \\ \hline
Valerio et al.                                                & 0.622             & --               & --                                                            \\ \hline
\begin{tabular}[c]{@{}c@{}}Karunakaran\\  et al.\end{tabular} & 0.47              & --               & 69.95\%                                                       \\ \hline
Our Model                                                     & 0.768             & \textbf{0.7543}  & \textbf{76.2\%}                                               \\ \hline
\end{tabular}
\end{table}
As can be seen, our model outperforms other models at Logloss and test accuracy. Also, our model gives the second best F1 score. We also tabulated the class-wise F1 scores in Table ~\ref{tab:f1scores}. For each class, F1-score $> 0.7$, showing that our solution performs significantly well in classifying the songs into all possible genre present in the dataset.

\begin{table}[thpb]
\caption{F1-scores for each class}
\label{tab:f1scores}
\centering
\begin{tabular}{|c|c|}
\hline
\textbf{Genre}    & \textbf{F1-score} \\ \hline
Rock              & 0.768             \\ \hline
Electronic        & 0.787             \\ \hline
Experimental      & 0.733             \\ \hline
Hip-Hop           & 0.716             \\ \hline
Folk              & 0.787             \\ \hline
Instrumental      & 0.731             \\ \hline
Pop               & 0.768             \\ \hline
International     & 0.713             \\ \hline
Classical         & 0.798             \\ \hline
Old-Time/Historic & 0.724             \\ \hline
Jazz              & 0.759             \\ \hline
Country           & 0.721             \\ \hline
Soul-RnB          & 0.711             \\ \hline
Spoken            & 0.723             \\ \hline
Blues             & 0.759             \\ \hline
Easy Listening    & 0.722             \\ \hline
\end{tabular}
\end{table}

\section{Conclusion}
\label{conc}
In this work we make use of two types of feature extractors in order to extract feature representations from multimodal data. Extracted feature vectors are fused in order to successfully predict the classes of corresponding music tracks. This fusion of features extracted using a HAN and a CNN results in a better performance as compared to models which consider only data of a single mode.

\bibliographystyle{unsrt}  


\end{document}